\input epsf
\documentstyle[preprint,eqsecnum,floats,aps]{revtex}
\def\D0{D\O}
\def\etmisv {\mbox{${\hbox{${\vec E}$\kern-0.6em\lower-.1ex\hbox{/}}}_T$}}
\def\etmis  {\mbox{${\hbox{$E$\kern-0.6em\lower-.1ex\hbox{/}}}_T$ }}

\def\ifmath#1{\relax\ifmmode #1\else $#1$\fi}%

\newcommand{\ttbar}     {\ifmath{\mathrm{t\bar{t}}}}

\begin{document}
\preprint{FNAL CONF-96-323}


\title{On Measuring the top quark mass using the dilepton decay modes
       \thanks{Submitted to the proceedings of the 1996 DPF/DPB study on new
        directions for High Energy Physics, Snowmass, Colorado.
        Work supported by DoE }}

\author{ Rajendran Raja\\
       {\it Fermi National Accelerator Laboratory } \\ 
       {\it P.O. Box 500 } \\
       {\it Batavia, IL 60510 }}

\maketitle

\begin{abstract}
We demonstrate a new likelihood method for extracting the top quark mass from  
events of the type 
$t\bar t \rightarrow bW^+(lepton+\nu) \bar bW^-(lepton+\nu)$. 
This method estimates the top quark mass correctly from an ensemble of 
dilepton events.  The method proposed by Dalitz and Goldstein \cite{dalitz}  is
shown to result in a systematic underestimation of the top quark mass. 
Effects due to the spin correlations between the top and anti-top
quarks are shown to be unimportant in estimating the mass of the top
quark.
\end{abstract}
\section{Introduction }

The \ttbar{} dilepton decay channels in which both the W's decay into leptons 
and neutrinos are under-constrained with respect to the the reconstruction of
the top quark mass due to the presence of the two missing neutrinos.
Nevertheless as Dalitz and Goldstein \cite{dalitz} and independently
Kondo\cite{kondo} et al have shown, it is possible to extract mass information
from these events using a likelihood method. For each event, solutions are
obtained for the kinematic quantities for a range of top quark masses. Each
solution is weighted by a product of structure functions which estimates the
probability of producing a \ttbar{} pair consistent with the event at that top
quark mass and a decay probability factor which neglects the polarization
of the top quark. In this paper we show that the Dalitz-Goldstein
weighting scheme leads to a systematic  underestimation of the top quark mass.
We propose a likelihood scheme which involves no kinematic weighting that is
shown to estimate the top quark mass correctly. Finally we show that {\em not}
allowing for the spin  correlations in  the decay of top quarks in the
Dalitz-Goldstein scheme does not further bias  the mass estimate significantly.

With the proposed luminosity upgrades of the Tevatron \cite{tev33}, it is
possible to acquire thousands of events of the type  $t\bar t \rightarrow
bW^+(lepton+\nu) \bar bW^-(lepton+\nu)$, where both the $b$ quark jets are
identified. The number of jet permutations in these channels is smaller than
the lepton + jets decay modes of the top quark. It may then become possible to
measure the top quark mass using the dilepton  channels with the least amount of
systematic error.

\section{Method}

Each dilepton event is characterized by 14 measurements, namely the three
vectors of the two $b$ jets, leptons and the missing $E_T$ vector of the event.
We denote these measurements collectively by the configuration vector $c$.
Kinematically, each event is characterized by 18 variables namely, the three
vectors of the $b$ jets, leptons and the two missing neutrinos. For any given
top quark mass, there are four constraints, that constrain the lepton and 
neutrino pairs to  the W mass and the W and $b$ pairs to the top quark mass.
Given a top quark mass, this enables one to solve for the neutrinos. This
results in a pair of quadratic equations for the transverse components of each
neutrino \cite{kondo}. The solution involves finding the intersection of two
ellipses. This can yield zero, two or four solutions for a given top quark
mass. The likelihood P($m|c$) of a solution  for a top quark mass $m$, given
the observed configuration vector  $c$, is obtained by using Bayes' theorem.

\begin{equation}
       P(m|c) = \frac{P(m)P(c|m)}{\int P(m)P(c|m)dm}
\end{equation}

Where P($m$) is the {\em a priori} probability distribution of the top quark
mass. 
P($c|m$) is the probability of observing the configuration vector $c$, for a
given top quark mass $m$. If after each event is analyzed, P($m$) is updated by
P($m|c$) iteratively, one gets the familiar multiplicative rule for combining
likelihoods. Dalitz and Goldstein \cite{dalitz,dalitz1} use the prescription
\begin{equation}
  P(c|m) = \Sigma_{partons} F(x_1)F(x_2)D(l_1,m)D(l_2,m)
\label{eqn1}
\end{equation}
where F($x_1$), F($x_2$) are  the probabilities of finding  partons with
momentum fraction $x_1$ and $x_2$ in the colliding beam particles consistent
with producing the event in question and D($l_1,m$) (D($l_2$,m)) is the
probability of observing a lepton of energy $l_1 (l_2)$ in the rest frame 
of the top (anti-top) quark. The expression for D($l,m$) as given in
\cite{dalitz} neglects the top quark polarization, but treats the subsequent
W decays according to the standard model.
In reality spin correlations are present and the two decays are correlated.
\subsection{Measurement errors}
 The expression for P($c|m$) in equation(\ref{eqn1}) must be further modified 
to take into account measurement errors. If the measured configuration vector
is $c_m$ of a true configuration vector $c$, we can write
\begin{equation}
   P(c_m|m) = \int P(c|m) R(c,c_m,\sigma) dc
\label{eqn2}
\end{equation}
where the function R($c,c_m,\sigma$) is the resolution function of the
experiment, denoting the probability of observing the configuration vector
$c_m$ given a true configuration vector $c$. The resolution of each of the
components of $c$ is contained in the resolution vector $\sigma$. In practice,
it is possible to choose the configuration vector $c$ such that R($c,c_m,
\sigma$)is Gaussian. Due to the symmetric nature of the Gaussian in $c$ and
$c_m$, we can
re-express equation( \ref{eqn2}) as
\begin{equation}
   P(c_m|m) = \int P(c|m) R(c_m,c,\sigma) dc
\end{equation}
This Gaussian integration can be carried out by smearing the measured
configuration $c_m$ repeatedly in a Gaussian fashion with standard deviations
$\sigma$ such that, for N smeared configurations,
\begin{equation}
   \frac{dN}{N} = R(c_m,c,\sigma) dc 
\end{equation}
The Monte Carlo integration then yields
\begin{equation}
   P(c_m|m) = \frac{1}{N} \Sigma_{configurations} P(c|m) 
\end{equation}

\subsection{Choice of the configuration vector }
In what follows, we will assume that both the leptons are electrons.   We choose
the three  quantities, energy, pseudo-rapidity and azimuth (E,$\eta,\phi$) to
define the three vectors of  the leptons and jets.  The electrons are smeared
with a typical collider detector fractional resolution of $15\%/\sqrt(E) $ in
energy and the jets with a fractional energy resolution of $80\%/\sqrt(E) \oplus
.05$. We ignore the fluctuations in direction, as these are dwarfed by the
energy fluctuations. The $p_T$ of the rest of the event after removing the
leptons and jets is also a measured quantity and is smeared as though it were a
small jet. The \etmisv is a deduced quantity from the measured quantities
listed. The case when one or both of the leptons is a muon is handled by
smearing the  inverse momentum of the muon as a Gaussian, but will not be
further discussed here.

We do not {\em a priori} know which lepton is associated with which $b$ quark.
We consider both combinations and add the likelihoods from either combination to
form the total likelihood for each event, which is normalized to unity when
integrated over the top quark mass $m$. 
\subsection{Combining likelihoods}

 We generate the likelihood spectrum for each event in the top quark mass
 range of 100- 250 GeV/c$^2$ at intervals of 1 GeV/c$ ^2$

The combined likelihood for an ensemble of events is obtained by multiplying the
likelihoods of the individual events. The likelihood for an individual event can
be zero for some values of the top quark mass due to the fact that we have used
a narrow resonance approximation for the W mass in finding the solutions, and
due to the finite number of smears  done per event. In order to prevent the
combined likelihood having zeroes in some bins due to these effects, we add a
uniform floor  probability distribution that integrates to  $1\%$, in the top
quark mass interval 100 - 250 GeV/c$^2$ to the likelihood distribution of each
event and renormalize it. The final mass values are insensitive to the exact
value of the floor.

 The individual event likelihoods are sampled at top quark mass intervals of 
1 GeV/c$^2$. The combined likelihood mass errors can fall below 1 GeV/c$^2$. We
interpolate the individual event likelihoods at mass intervals of 
0.25 GeV/c$^2$ so that the final combined event likelihood can span several bins
in mass.

 In general Monte Carlo events have weights associated with them. These were
normalized so that the average weight in the event sample was unity. Events with
weights outside the window 0.3- 3.0 were rejected. The likelihood distribution
for each event was raised to the power given by its weight before being used to
form the combined likelihood.
\subsection{Event selection criteria}
 We select only those events with $E_T >$ 15 GeV for both the
 leptons and jets and $\etmis{} >25 GeV$. We demand that both the $b$ jets are
 explicitly identified by a tagging algorithm. While smearing, we only admit
 smeared configurations that satisfy the same criteria as the event selection. 

 In what follows we smear each Monte Carlo generated event once to simulate the
 measurement process and subsequently 1000 times to do the Monte Carlo
 integration.
\section{Results}
We generate Monte Carlo events with a top quark mass of 175 GeV/c$^2$. We
neglect top quark polarization in generating these events, but treat the
subsequent W decays according to the standard model\cite{parke}. 
No final state or initial state radiation is included in this
initial set of events.
The events have \ttbar{} pairs produced
according to the standard QCD  processes (dominated at Fermilab energies by
valence quark fusion and $s$ channel gluon exchange). The top quark polarization
is neglected after production. The W's are decayed correctly according to the
standard model, mimicking the assumptions going into the Dalitz-Goldstein
weighting scheme. We call this the uncorrelated sample.

\begin{figure}[h]
\epsfxsize = 16.cm
\epsffile{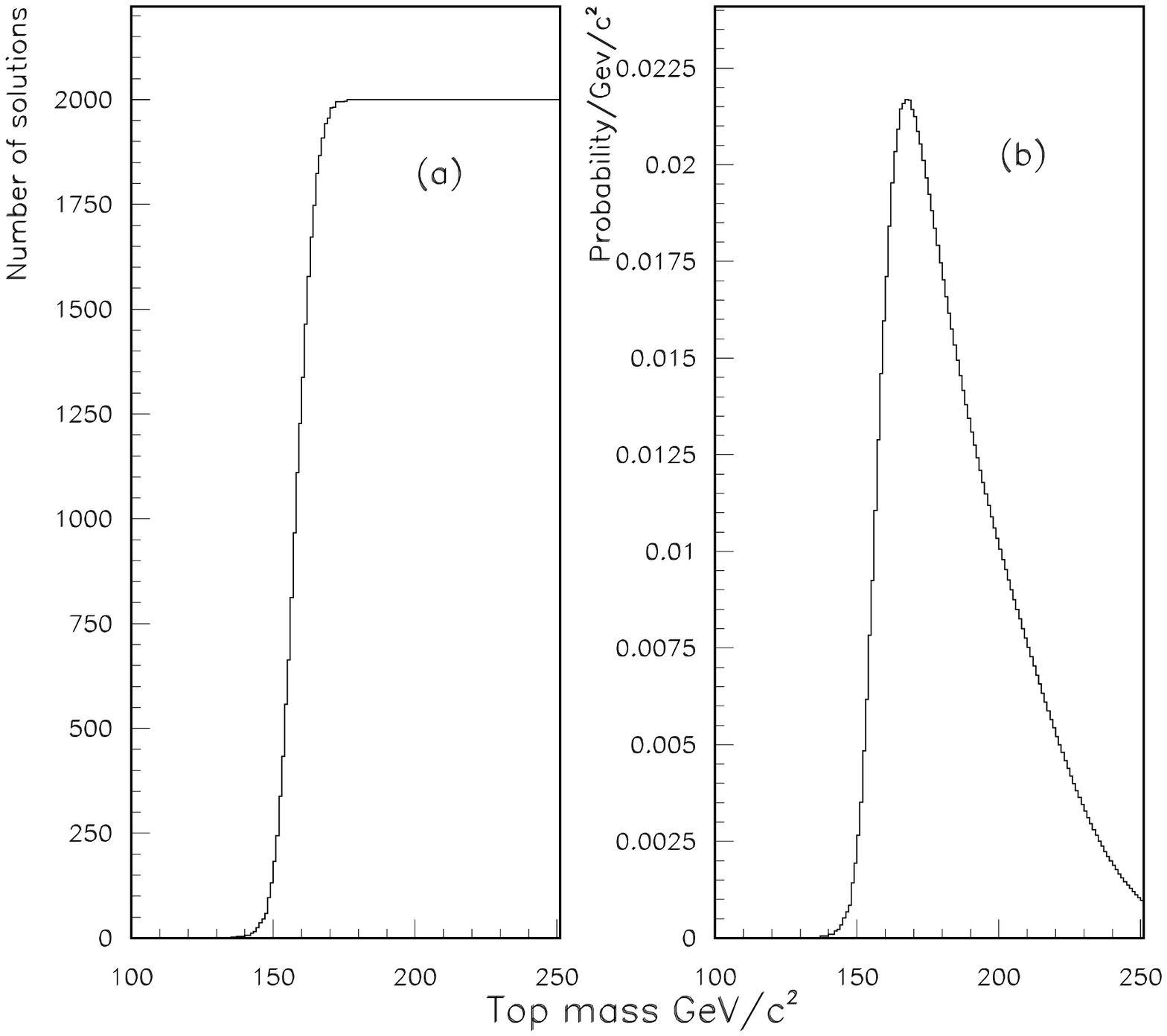}
\caption{(a) shows the number of solutions versus top quark mass for a typical
event generated with top quark mass of 175 GeV/c$^2$. (b) Probability
distribution for that event obtained according to the Dalitz-Goldstein
prescription.}
\label{fig1}
\end{figure}

\begin{figure}
\epsfxsize = 16.cm
\epsffile{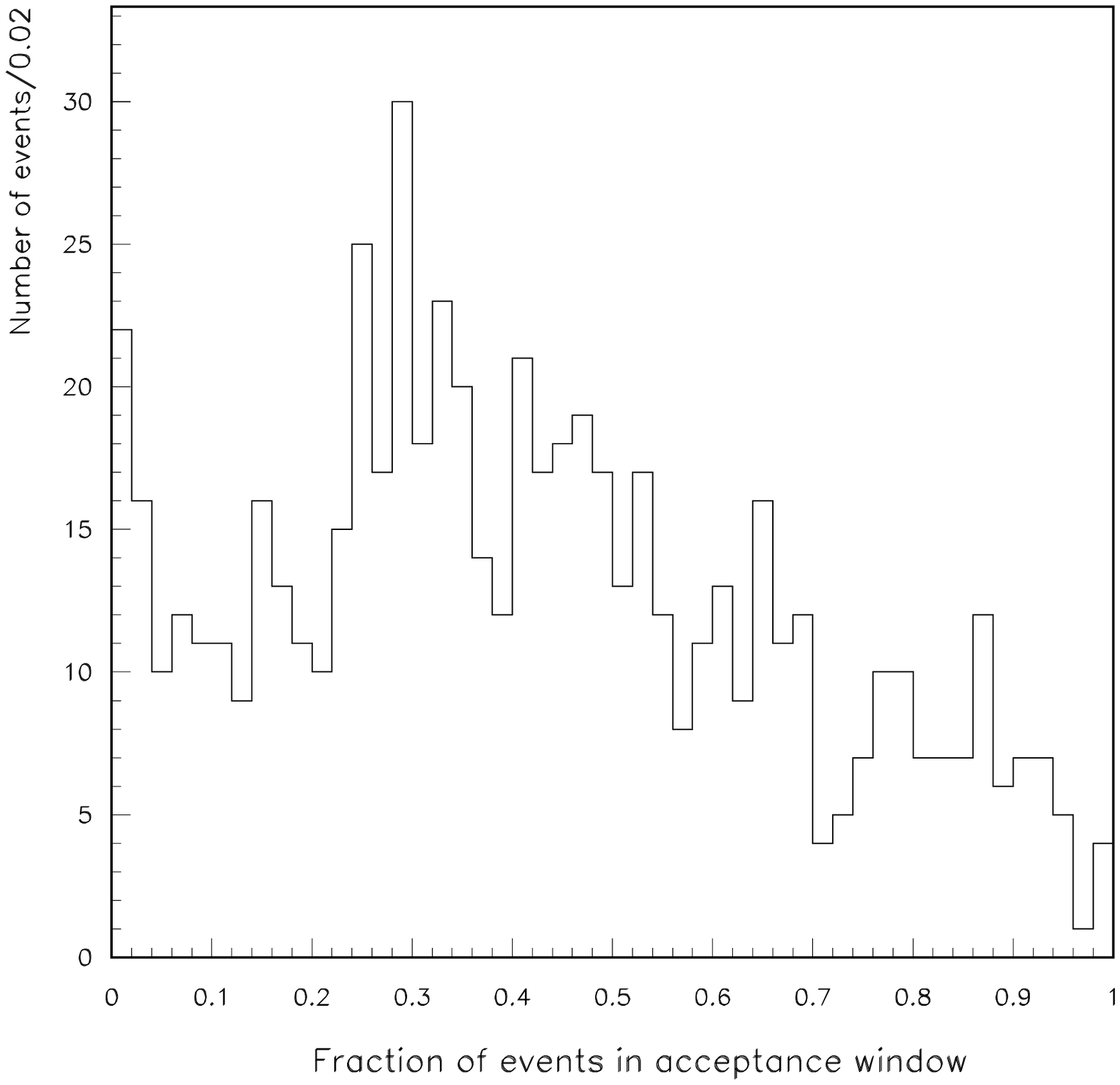}
\caption{Histogram of the fraction of the number of solutions $\cal R$ in a
window  $\pm$ 35 GeV/c$^2$ of the generated top quark mass.}
\label{fig2}
\end{figure}

Figure (\ref{fig1}(a)) shows the unweighted distribution of solutions found for
the $\approx$ 1000 smeared configurations for a typical such event. The
solutions turn on at a mass of 140 GeV/c$^2$ and stay turned on till the end of
the mass range at 250 GeV/c$^2$. Figure (\ref{fig1}(b)) shows the probability
distribution for this event using the Dalitz-Goldstein prescription of equation
(\ref{eqn1}). The structure function weighting in equation (\ref{eqn1}) makes
the high mass solutions less likely yielding a likelihood distribution that has
a distinct peak.
We now proceed to analyze a sample of $\approx$ 1000 such Monte Carlo events 
that  decay into dileptons.
Because of measurement errors, not all of these events will give
solutions consistent with a top quark in the mass range 100-250 GeV/c$^2$.
 Figure (\ref{fig2}) is a histogram of the quantity  $\cal R$  defined by
\begin{equation}
{\cal R} = \Sigma_{window}  \frac{N_i}{tot_M \times  N_{smear}}
\end{equation}
where $N_i$ is the number of solutions for top quark mass $i$ , $tot_M$ is the
total number of top quark masses considered and $N_{smear}$ is the total number
of smears per event. The sum extends for top quark masses in a window $\pm$ 
35 GeV/c$^2$ of the generated top quark mass. There is a peak in the histogram
for values of $\cal R$ below 0.1. This is due to events that are so mismeasured 
that they have difficulty solving for a top quark mass in the window considered
even when smeared a thousand times. We reject events with $\cal R < $0.2 
since these will have very spiky likelihood distributions.
Figure (\ref{fig3}(a)) is the combined likelihood of 511 events  which  survive
after event selection criteria and the $\cal R$ cut from an initial sample of
925 events, using the Dalitz-Goldstein weighting scheme  \cite{struc}.  

The most likely top quark mass from the event sample is  164.5 $\pm$ 0.54
GeV/c$^2$. The Dalitz-Goldstein weighting scheme thus introduces a bias of  
10.5 GeV/c$^2$ towards lower masses at this value of the top quark mass.
\subsection{A Critique of the Dalitz-Goldstein weighting scheme}
For a given event, the parton momenta ($x_1,x_2$) needed to produce it will
decrease as the top quark mass $m$ is decreased since $x_1 x_2 = m^2/s$, where
$s$ is the overall center of mass energy squared. This means that the
Dalitz-Goldstein weighting scheme will tend to skew the likelihood distribution
for each event toward lower top quark masses, since it is proportional to the
product of the structure functions. We note that the top quark production cross
section is also a product of such structure functions and decreases rapidly as
the top quark mass increases, for the same reason. The likelihood scheme
proposed by Kondo et al \cite{kondo} is proportional to the top quark production
cross section and also suffers from this defect. It is this skewing of the
likelihood distributions towards lower masses that produces a 10.5 GeV/c$^2$
bias in the Dalitz-Goldstein scheme. One can indeed ask why the top quark mass
measurement has to be coupled to its production mechanism at all.
\subsection {A new likelihood method}

\begin{figure}
\epsfxsize = 16.cm
\epsffile{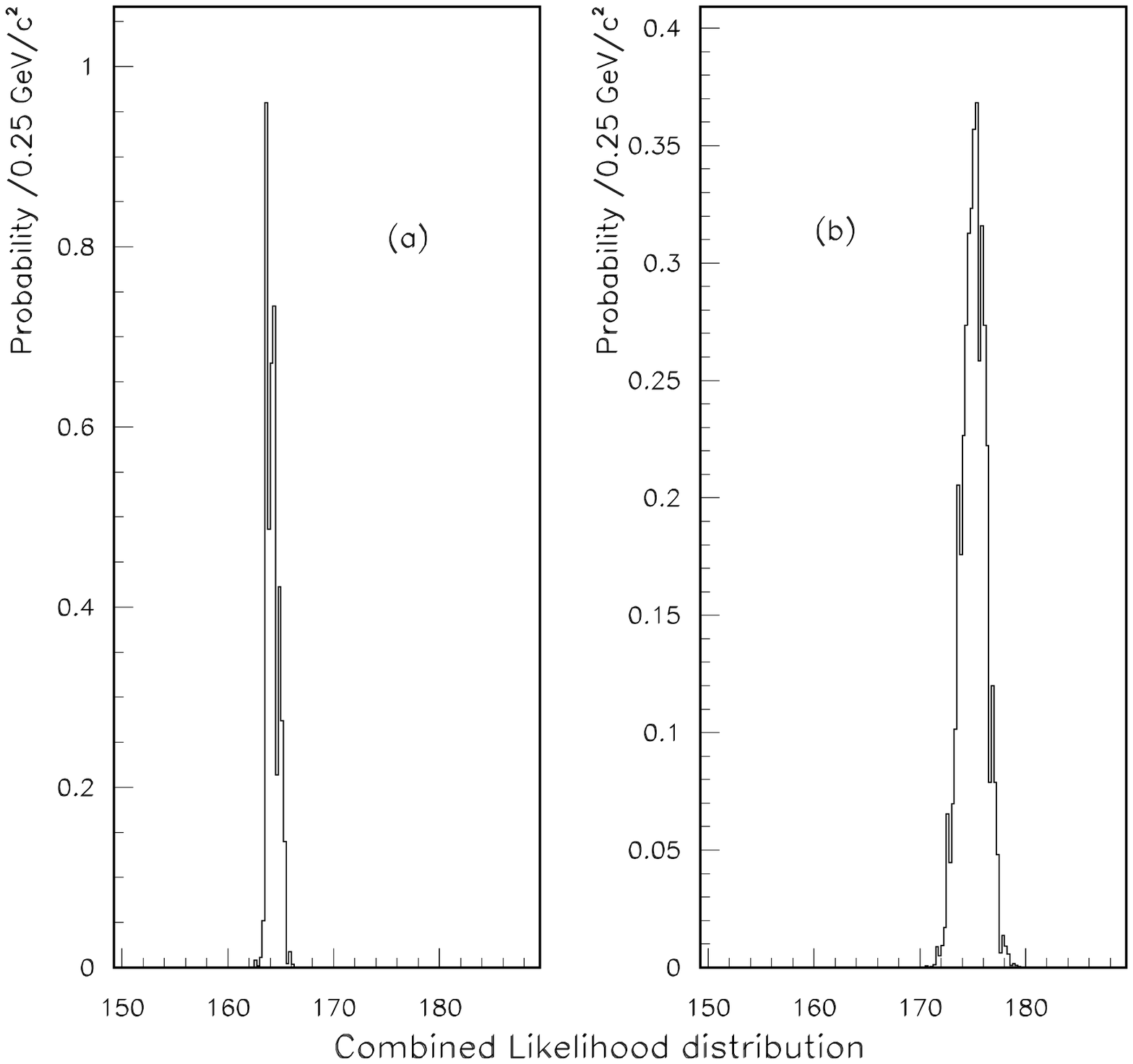}
\caption{For events generated with a top quark mass 175 GeV/c$^2$,
(a)Combined likelihood distribution using  the 
Dalitz-Goldstein weighting
scheme, yields a mean top quark mass  164.5 GeV/c$^2 \pm $  0.5 GeV/c$^2$.
(b) using the new likelihood method proposed here, yields a mean top quark
mass 175.3 GeV/c$^2 \pm$ 1.1 GeV/c$^2$. }
\label{fig3}
\end{figure}

\begin{figure}
\epsfxsize = 16.cm
\epsffile{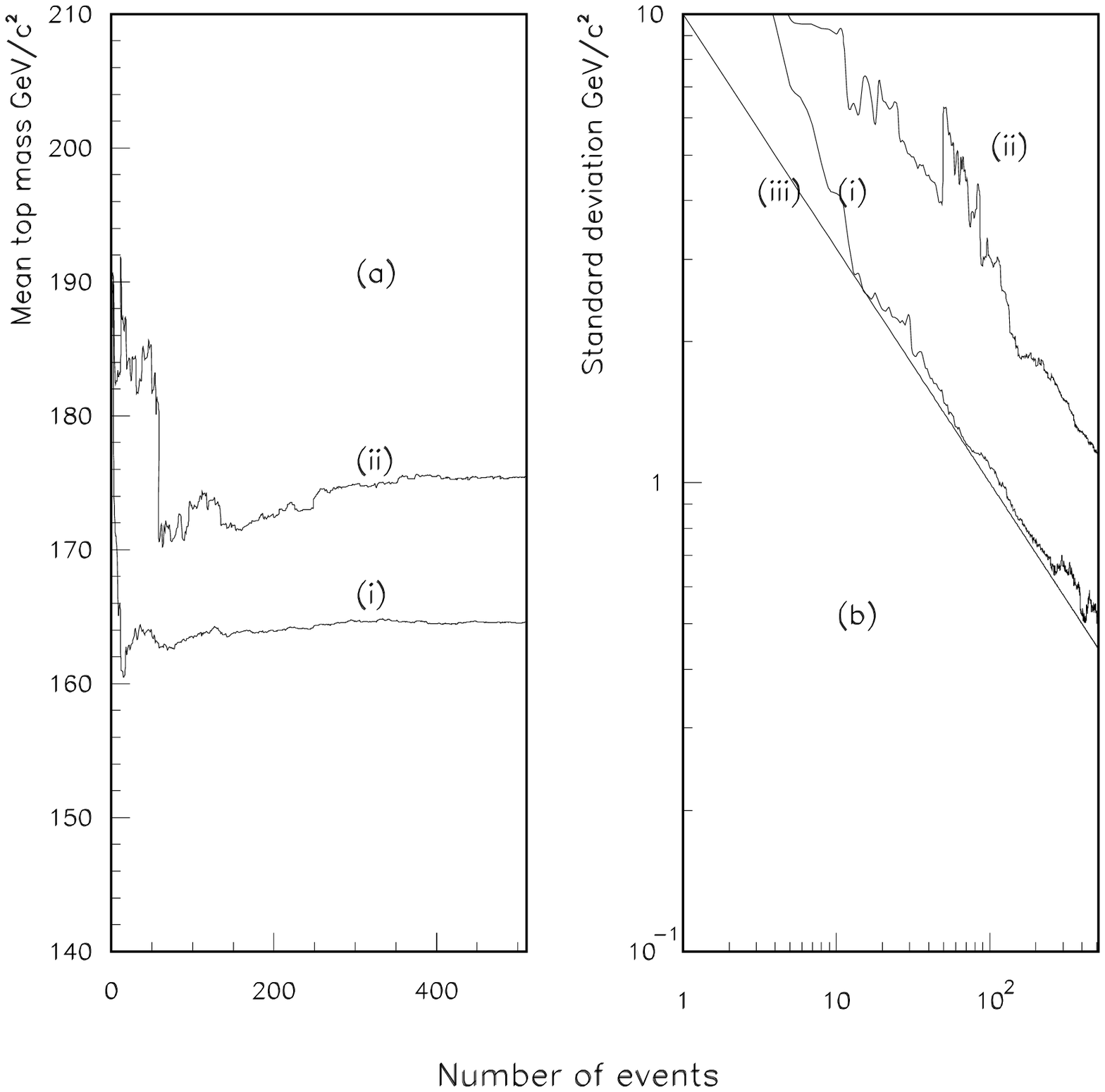}
\caption {Evolution of (a) the mean value (b) standard deviation 
of the combined likelihood distribution as a function of the 
number of events for (i) Dalitz-Goldstein weighting scheme, 
(ii) The ``no-weights" method,
(iii) Curve showing $N^{-1/2}$  shape}
\label{fig4}
\end{figure}

Figure (\ref{fig1}(a)) shows the number of solutions for a typical event as a
function of the top quark mass. We now make the radical proposal of not using
any weights at all, but simply use a likelihood distribution that is shaped like
the number of solutions as a function of the top quark mass. If one examines
this distribution visually for an ensemble of top quark events, there exist
a significant number of events where the likelihood distribution thus formed
does show a peak and falls for large top quark masses. Using this scheme, one
gets the combined likelihood of Figure (\ref{fig3}(b)) which peaks at the input
mass, but has a larger standard deviation. The larger standard deviation is due
to the fact that we are not suppressing the high mass tail of the individual
event likelihood distributions using a weighting scheme. This method does not
use any extrinsic information of the top quark production mechanism to obtain
the mass but relies solely on the measured kinematic quantities of the events in
question. We christen this scheme the ``no-weights" method. 
Figure (\ref{fig4}(a)) shows the evolution of the mean value of the combined
likelihoods for the Dalitz-Goldstein method and the no-weights method as a
function of the number of events. Figure (\ref{fig4}(b)) shows the evolution of
the standard deviation \cite{std} of the combined likelihoods using the two
methods as a function of the number of events. An approximate $1/\sqrt(N)$
dependence on the number of events is evident.
The ``no-weights" mass is slightly sensitive to the value of the 
$\cal R$ cut, since the events rejected by the $\cal R$ cut tend to favor
lower top masses. It is possible to adjust the $\cal R$ cut so that the input
top mass is returned by the ``no-weights" algorithm. Once tuned at one generated
top quark mass, the algorithm works well at all other masses with the cut
unchanged. The Dalitz-Goldstein scheme cannot reproduce the generated  mass for
any value of the $\cal R$ cut. It should be noted that the  window chosen around
the generated mass in defining the $\cal R $ cut  has to be symmetric about the
generated mass  to avoid bias. This can be done iteratively when dealing with
data.
\begin{table}[t]
\begin{center}
{\footnotesize 
\begin{tabular}{|c||c|c|} \hline\
Top mass   & Dalitz-Goldstein & No-weights \\
MC  sample & method GeV/c$^2$ & method GeV/c$^2$ \\
\hline
175 GeV/c$^2$  & 164.5 $\pm$ 0.54 & 175.3 $\pm$ 1.11 \\   
Spin uncorrelated &               &                  \\
\hline
175 GeV/c$^2$   & 164.8 $\pm$ 0.49  & 174.1 $\pm$ 1.05 \\   
Spin correlated   &                 &                  \\
\hline
\hline
140 GeV/c$^2$& 131.8 $\pm$ 0.37  & 139.9 $\pm$ 0.7 \\   
Isajet       &                   &                  \\
\hline
160 GeV/c$^2$& 147.6 $\pm$ 0.48  & 158.0 $\pm$ 1.02 \\   
Isajet       &                   &                  \\
\hline
180 GeV/c$^2$& 163.7 $\pm$ 0.74  & 175.1 $\pm$ 0.92 \\   
Isajet       &                   &                  \\
\hline
200 GeV/c$^2$& 179.7 $\pm$ 0.58  & 193.2 $\pm$ 1.08 \\   
Isajet       &                   &                  \\

&&\\    \hline\hline
\end{tabular}
}
\end{center}
\caption[]{Summary of top quark mass measurements on various 
Monte Carlo samples}
\label{tab1}
\end{table}
\subsection{Spin correlations and final state radiation effects}
We now generate events where both the top and anti-top quark polarizations are
taken into account and all spin correlations are kept at the tree level 
\cite{kleiss}.
We use the two weighting methods outlined above to determine the
top quark mass. The results are presented in table (\ref{tab1}). There is no 
apparent shift in the top quark mass between the two samples for either method.
From this, we conclude that spin correlations do not affect  the
determination of the top quark mass in the dilepton channel in any significant
way. 
The Monte Carlo samples used so far do not include additional jets due to
initial and final state gluon radiation. We now generate $\approx$ 1000
events at top quark masses of 140,160,180 and 200 GeV/c $^2$ using the program
Isajet \cite{isa}. We demand that both the $b$ quark jets are identified. Table
(\ref{tab1}) shows the results using either method. Once again, the
Dalitz-Goldstein method underestimates the generated mass. The ``no-weights"
method can now be used to estimate the effects due to final state radiation
as implemented in Isajet. It can be seen that the net effect of the 
final state radiation is to systematically lower the measured value of the top
quark mass. The amount of lowering increases with the top quark mass, due to the
increased amount of final state radiation. At a top quark mass of 180 GeV/c$^2$,
the effect of final state radiation is to lower the top quark mass by 
$\approx$ 5 GeV/c$^2$.
 Finally, we have also studied the effect of the event selection $E_T$ cuts
for their effect on the  result. We get results that are the same within errors,
even when no $E_T$ cuts are used.
\section{A proposal for a correct weighting scheme}
 If one insists on weighting events using production and decay information from
 the standard model, the expression for P($c|m$) has to have the following
 properties.
\begin{equation}
   \int P(c|m) dc = 1
\end{equation}
An expression that satisfies this is given by
\begin{equation}
    P(c|m) = \frac{1}{\sigma_{vis}(m)} \frac{d\sigma_{vis}(m)}{dc}
\end{equation}
where $\sigma_{vis}(m)$ is the top quark production {\em visible} in the 
detector. The biasing effect in the top quark mass due to the structure function
product is removed by division by the function  $\sigma_{vis}(m)$.
The configuration vector can be chosen as any set of measured variables, since
the resulting expression for P($m|c$) is invariant under a change of variables
\cite{jacob}. However, equation (\ref{eqn2}) , implies a unique set of variables
for the configuration vector $c$, since these are the quantities that are
fluctuated in a Gaussian fashion. We will report on results using this weighting
scheme in a forthcoming paper.
\section{Conclusions}
 We have demonstrated a new likelihood method that determines the top quark mass
in dilepton decays of the top quark that gives an  unbiased estimate of the top
quark mass. We demonstrate that weighting schemes that involve products of
structure functions such as the Dalitz-Goldstein scheme, give a downward bias to
the measured value of the top quark mass. We demonstrate that spin correlation
effects between the top and anti-top decay products do not influence the 
outcome of the mass measurement. We estimate the effects due to final state
radiation as implemented in Isajet.

The statistical precision obtainable using a thousand top to
dilepton fully tagged events using this method is of the order of a GeV/c$^2$
using this technique. Assuming that jet energy scale systematics in the upgraded
Tevatron detectors can be controlled to this level, the dilepton channels
provide an excellent means of measuring the top quark mass.

\section{Acknowledgements}
 The author wishes to thank Stephen Parke for helpful discussions and for 
providing the non-Isajet  top Monte Carlo samples.


\begin{thebibliography}{99}
\bibitem{dalitz1} Strictly the method outlined here is an extension of the
method  used in \cite{dalitz} since they were working on one CDF event for
which the \etmis{} was not known and as such had two additional missing
variables.
\bibitem{struc}We use EHLQ set I structure functions for this calculation.
The results are largely insensitive to the structure function type.
\bibitem{std} The standard deviation of the likelihood is an approximate 
measure of the error in the top quark mass, which becomes more accurate as the
likelihood distribution becomes more Gaussian. We use it here, since it 
is readily calculated.
\bibitem{jacob} The Jacobian of the transformation between the different set of
measured variables is independent of the top quark mass for any given event and
gets divided out while normalizing the likelihood P($m|c$) to unity.
\end{thebibliography}
\end{document}